# Distributed Bilevel Energy Allocation Mechanism with Grid Constraints and Hidden User Information

M.Nazif Faqiry, *IEEE Member*, Sanjoy Das

Citation (APA)

# Distributed Bilevel Energy Allocation Mechanism with Grid Constraints and Hidden User Information

M.Nazif Faqiry, *IEEE Member*, Sanjoy Das

*Abstract*— A novel distributed energy allocation mechanism for distribution system operator (DSO) market through a bilevel iterative auction is proposed. With the locational marginal price known at the substation node, the DSO runs an upper level auction with aggregators as intermediate agents competing for energy. This DSO level auction takes into account physical grid constraints such as line flows, transformer capacities and node voltage limits. This auction mechanism is a straightforward implementation of projected gradient descent on the social welfare (SW) of all home level agents. Aggregators, which serve home level agents - both buyers and sellers, implement lower level auctions in parallel, through proportional allocation of power, until market equilibrium is established. This is accomplished without asking for the agents' utility functions and generation capacities that are considered private information. The overall bilevel auction is shown to be globally efficient and weakly budget balanced.

*Index Terms*—distribution system; agents; trading; auction; bid; social welfare

## NOMENCLATURE

| | |
|---|---|
| $\mathcal{N}$ | Set of nodes, excluding root |
| $N$ | Cardinality $|\mathcal{N}|$ of $\mathcal{N}$ |
| $\mathcal{A}$ | Set of nodes with aggregators |
| $A$ | Cardinality of $\mathcal{A}$ |
| $\delta$ | Maximum allowable per unit (pu) voltage deviation |
| $P_0, Q_0$ | Active, Reactive power from root node in pu |
| $V_0$ | Voltage at root in pu |
| $k, l$ | The $k^{\text{th}}$, $l^{\text{th}}$ nodes, $k, l \in \mathcal{N}$ |
| $\mathcal{D}(k)$ | Set of downstream nodes of $k$ (immediate & separated) |
| $u(k)$ | Index of immediate upstream node of node $k$ |
| $\mathcal{U}(k)$ | Index of all upstream nodes of $k$ to root, $k \in \mathcal{U}(k)$ |
| $r_k, x_k$ | Resistance and reactance of line $(u(k), k)$ in pu |
| $p_k$ | Real power injected into $k$ in pu, $k \notin \mathcal{A} \Rightarrow p_k = 0$ |
| $q_k$ | Reactive power injected into $k$ in pu, $k \notin \mathcal{A} \Rightarrow q_k = 0$ |
| $P_k$ | Active power flowing through line $(u(k), k)$ in pu |
| $Q_k$ | Reactive power flowing through line $(u(k), k)$ in pu |
| $\Delta V_k$ | Voltage drop through line $(u(k), k)$ in pu |
| $V_k$ | Voltage at node $k$ in pu |
| $\bar{S}_k$ | MVA limit of line going into node $k$ ( line $k$) in pu |
| $\theta_k$ | Fraction of $p_k$ as $q_k$ at node $k$ when $k \in \mathcal{A}$ |
| $c_k$ | Per unit price of energy of node $k$ in (¢/pu), $k \in \mathcal{A}$ |
| $\mathcal{N}_B^k$ | Set of buyers at node $k$, $k \in \mathcal{A}$ |
| $\mathcal{N}_S^k$ | Set of sellers at node $k$, $k \in \mathcal{A}$ |
| $N_S^k$ | Cardinality of $\mathcal{N}_S^k$ |
| $N_B^k$ | Cardinality of $\mathcal{N}_B^k$ |
| $i, j$ | The $i^{\text{th}}$ buyer and $j^{\text{th}}$ seller |
| $g_j^k$ | Max generation of $j^{\text{th}}$ seller in pu at node $k$ |
| $d_i^k$ | Demand in pu delivered to the $i^{\text{th}}$ buyer at node $k$ |
| $s_j^k$ | Supply in pu by to the $j^{\text{th}}$ seller at node $k$ |
| $b_i^k$ | Total bid money for demand $d_i^k$ by the $i^{\text{th}}$ buyer |
| $u_i^k, v_j^k$ | Utility functions of buyer $i$ and seller $j$ at node $k$ |
| $c_0$ | Price function at root (substation) node in ¢/pu |
| $\Theta_k$ | Social welfare function of aggregator $k$ |
| $\Omega$ | Global (DSO level) social welfare function |

## I. INTRODUCTION

THE proliferation of renewable energy resources (RES) at the distribution level is reshaping the market structure of distribution system operators (DSO). The electricity sector has devolved from a highly regulated system operated by vertically integrated utilities to a relatively decentralized system based more fully on market valuation and allocation mechanisms [1]. A RES owner with a surplus of energy is anticipated to participate in such mechanisms more strategically while seeking profit [2], [3]. Specifically, it sells energy with the objective of maximizing its payoff, i.e. the sum of the monetary gain from supplying an amount of energy and utility it gains from retaining any surplus energy that is not traded. In a similar manner, the payoff of an energy consumer, i.e. a buyer, is the difference between its utility gained from consuming energy and the cost of procuring that energy.

DSOs on the other hand, are expected to leverage the available local resources in order to capture additional value by optimizing the system for least cost operation while maintaining the physical system operation constraints [4]. One of the key challenges for efficient energy distribution mechanisms is its design to motivate active participation of customers [5]. Without their participation, the benefits of smart grid is not fully realized [6]. Therefore, suitable mechanisms for it to operate effectively within its physical constraints, while incentivizing customer participation, are needed.

The existing literature on the energy grid focuses mostly on pricing based on the distribution locational marginal price (DLMP) that is defined as the marginal cost of serving the next increment of the electrical demand at a distribution node. DLMP is typically determined in a centralized fashion by the Lagrange multiplier of the distribution node energy balance constraint [7]-[12]. It has been shown in [13] that introducing price responsive customers causes distribution line congestion. In [9], DLMP based pricing is used as the means of dynamic pricing tariffs to alleviate distribution system congestion.

This work was partially supported by the National Science Foundation-CPS under Grant CNS-1544705.

Similar to [8], [9], the recent work in [10] proposes two benchmark pricing methodologies, namely DLMP and iterative DLMP (iDLMP), for a congestion free energy management by buildings providing flexible demand. Aggregators in this model have contracts with flexible buildings to decide their reserve and energy schedule by interacting with the DSO in a cost optimal manner in order to avoid any congestion in the distribution grid. Standard dual decomposition [14] is used to determine iDLMP at the DSO level to alleviate the need for data transfer among the DSO and aggregators. This study does not simulate market conditions. Its lower level agents, i.e. buildings, do not have to place bids in order to obtain energy, which is supplied as per contractual obligations with the aggregator.

In this paper, a novel bilevel distribution auction mechanism is proposed that converges to a socially optimal solution, while maintaining physical grid constraints. The lower level auction, referred to as the aggregator level auction (ALA), is conducted by the local aggregators assigned to each distribution node among downstream consumers and prosumers.

The upper level auction, referred to as the DSO level auction (DLA), is implemented iteratively by the DSO among aggregators competing for the share of energy that the DSO receives from the wholesale market. The goal of the DSO's auction is draw a suitable amount of energy from the wholesale market and allocate it among competing aggregators in such a manner that maximizes the global SW, while maintaining physical grid constraints such as voltage, line, and transformer limits.

There are several aspects of the proposed auction mechanism that deviate from earlier research. Firstly, consistent with economic theory, equilibrium is established by means of a *market-driven* mechanism, where both the volume of energy traded as well as the equilibrium price, are determined entirely through the ALA bidding process. Thus, determining the energy available to each aggregator is shifted to the upper level DLA. Therefore, the underlying optimization algorithm is parallelized through primal (instead of dual) decomposition.

Next, the proposed bilevel mechanism is *efficient* – it maximizes the social welfare (SW) of the grid, which is the sum total of the utilities of the consumer and prosumer agents. This is different from revenue-oriented (or *optimal*) auctions, which try to maximize the revenue accrued by the DSO or aggregators [15]-[17]. The underlying optimization formulation in [10] which minimizes the total cost of procuring energy, is another example of an optimal mechanism. Efficient mechanisms that have been proposed so far for the energy grid, are *indivisible* (i.e. discrete) resource auctions, designed for load scheduling [18], [19], whereas this research treats energy as a divisible (i.e. continuous) resource.

*Information privacy* is a significant advantage of the proposed mechanism, where the agents reveal neither their utility functions nor their individual PV-generated energy to the aggregators. This is possible as at market equilibrium, SW is automatically maximized separately within each aggregator, without the need for any specific optimization algorithm.

Lastly, *asymmetric bidding* [3] is another feature of the proposed mechanism. Buying agents declare monetary amounts to the aggregator and are allocated energy. On the other hand, selling agents receive unit costs of energy from the aggregator and submit to the latter, the quantity of energy they are willing to sell. Although it is possible for an agent to switch roles at any stage of the mechanism, in doing so, it would have to change its bidding strategy accordingly. For simplicity, in the simulations reported here (Section III), it is assumed that agents declare their intention to act as buyers or sellers *a priori*. Bidding asymmetry is the means through which proportional fairness can be incorporated in a double-sided market-driven auction mechanism [3].

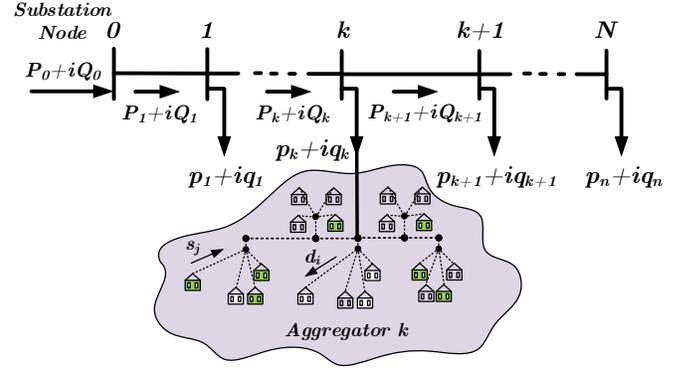

Fig. 1. Schematic diagram of a unidirectional single branch radial distribution system with N nodes excluding the root node.

The buying agents are allocated power directly in proportion to the monetary bids that they place, whereas the sellers are reimbursed monetary amounts that are proportional to the energy sold. Although such proportionally fair auctions possess certain desirable properties [40], these mechanisms are susceptible to price anticipation, where each agent bids in a manner to influence the equilibrium market price in its favor. Unfortunately, when all agents become price anticipators, the overall social welfare decreases. A novel approach to avoid price anticipation, through virtual bidding, has been addressed in details in [3].

## II. FRAMEWORK

Consider the radial distribution network in Fig. 1, with $N$ nodes excluding the root. The proposed bilevel auction mechanism is implemented in two levels among aggregators assigned to primary distribution nodes through DLA, and among strategic consumers and prosumers, i.e. agents residing on a lateral feeder, by the aggregators through ALA. In ALA, each agent's objective is to maximize its own profit by participating in its ALA. Each aggregator's objective is to maximize its agents' SW, without access to their private utility functions and generation capacities, which are shown in Fig. 2 with dotted lines and boxes. The aggregators make this possible by using double-sided proportional allocation [3] and participate in DLA by competing with other aggregators in order to get their optimal demand/supply share of the real power $p_k$. The DSO's objective is to implement DLA iteratively until convergence, so that maximum global SW is attained.

As shown in Fig. 2, at each iteration of DLA, aggregator $k$ receives real power $p_k$, implements ALA and submits its per unit price $c_k$ to trade $p_k$. The prices $c_k$ are market equilibrium price of multiple parallel ALA algorithms. Upon receiving updated $c_k$ from each aggregator $k \in \mathcal{A}$, the DSO executes the next iteration of DLA to find the new supply $p_k$. This procedure continues until convergence is reached. From the flow of information in Fig. 2, it is clear that the buying and selling agents do not reveal their utilities to the aggregator. Similarly, the agents' generations are not disclosed either.

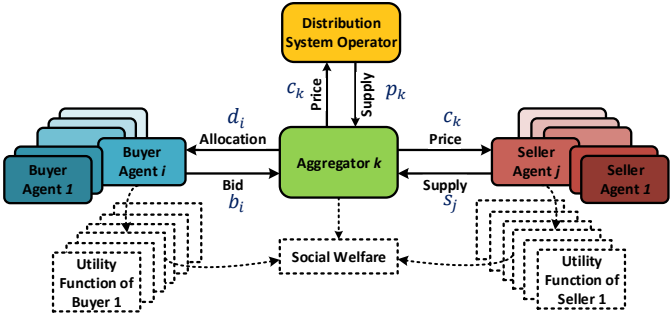

Fig. 2. Schematic showing flow of bidding information among agents in the bilevel mechanism.

*A. Distribution System Constraints*

The DSO needs to ensure that grid constraints are not violated. It is assumed in this research that the distribution system is balanced and all quantities can be represented per phase at each seller or buyer node. The simplified version of DistFlow equations [20], [21] that has been extensively used in literature [22]-[24] is adopted here. These equations are as shown below,

$$P_k = p_k + \sum_{l \in \mathcal{D}(k)} p_l, \quad (1)$$

$$Q_k = q_k + \sum_{l \in \mathcal{D}(k)} q_l, \quad (2)$$

$$V_k = V_0 - \sum_{l \in \mathcal{U}(k)} \Delta V_l, \quad (3)$$

where $\Delta V_l = \frac{r_l P_l + x_l Q_l}{V_0}$. $V_0$ is the root node's pu voltage and $\Delta V_l$ is the voltage drop at line segment entering node $k$. $\mathcal{D}(k)$ is the set of immediate and separated downstream nodes of $k$ and $\mathcal{U}(k)$ is the set of upstream nodes (including $k$) to the root node. For instance, in Fig. 3, $\mathcal{D}(17) = \{18, 19, 20, 21\}$, $\mathcal{U}(27) = \{1,2,27\}$. The following system architecture matrices are defined,

$$[\mathbf{A}]_{kl} = \begin{cases} 1 & k = \text{aggregator } l \\ 0 & \text{otherwise,} \end{cases}$$

$$[\mathbf{D}]_{kl} = \begin{cases} 1 & l \in \mathcal{D}(k) \text{ or } k = l \\ 0 & \text{otherwise,} \end{cases}$$

$$[\mathbf{U}]_{kl} = \begin{cases} 1 & l \in \mathcal{U}(k) \\ 0 & \text{otherwise.} \end{cases}$$

In the above, $\mathbf{A}$ is an $N \times A$ matrix and $\mathbf{D}, \mathbf{U}$ are $N \times N$ matrices associated with the spatial topology of the radial distribution network. The matrix $\mathbf{A}$ is the node-aggregator matrix that has an entry of unity ('1') at every column on the row, i.e. node, with which it is associated. The matrices $\mathbf{D}$ and $\mathbf{U}$ correspond to the descendant and ancestor nodes. Every row (node) of $\mathbf{D}$ and $\mathbf{U}$ has an entry of unity where the corresponding column (node) is its descendant or ancestor, and a zero elsewhere. The voltage drop in line $k$ is $\Delta V_k = r_k P_k + x_k Q_k$, so that,

$$\Delta \mathbf{V} = V_0^{-1} (\mathbf{r} \circ \mathbf{P} + \mathbf{x} \circ \mathbf{Q}) \quad (4)$$

where $\circ$ is the elementwise (Hadamard) product and $\Delta \mathbf{V}$ is the $N \times 1$ vector whose $k^{\text{th}}$ entry is $\Delta V_k$. This notation has been followed throughout the remainder of this paper – bolded variables are vectors of their italicized scalar counterparts, with the entry position specified in the latters' subscripts.

Using the matrices defined above, (1) – (3) can be written as follow,

$$\mathbf{P} = \mathbf{DAp}, \quad (5)$$

$$\mathbf{Q} = \mathbf{DAq}, \quad (6)$$

$$\mathbf{V} = V_0 \mathbf{1}_N - \mathbf{U} \Delta \mathbf{V}. \quad (7)$$

Note that it is assumed that homes are furnished with smart meters, and in case of sellers, also inverters [25], [26]. They are also capable of communicating their reactive power supply/demand to/from the aggregators at the end of each ALA.

In order to enforce the physical grid constraints, the DSO requires the reactive powers $q_k$ of the aggregators in addition to their $p_k$s, with the latter directly available through the bidding process in ALA. The ratio between $q_k$ and $p_k$ is denoted as $\theta_k$, so that $q_k = \theta_k p_k$. Each aggregator $k$ can obtain $q_k$ as the sum of the reactive power requirement obtained directly from its agents in $\mathcal{N}_B^k \cup \mathcal{N}_S^k$ at the termination of each round of ALA. Alternately, $q_k$ can be estimated from historical data. In this research, it is assumed that the aggregator has access to the numerical value of $\theta_k$, that is communicated to the DSO along with $c_k$. With $\boldsymbol{\theta} = [\theta_k]_{k \in \mathcal{A}}$ and $\mathbf{q} = \boldsymbol{\theta} \circ \mathbf{p}$ as elementwise product of $\boldsymbol{\theta}$ and $\mathbf{p}$, using (4), the expressions in (5) – (7) can be rewritten as,

$$\mathbf{P} = \mathbf{DAp}, \quad (8)$$

$$\mathbf{Q} = \mathbf{DA}(\boldsymbol{\theta} \circ \mathbf{p}), \quad (9)$$

$$\mathbf{V} = V_0 \mathbf{1}_N - \frac{1}{V_0} \mathbf{U}(\mathbf{r} \circ \mathbf{DAp} + \mathbf{x} \circ \mathbf{DA}(\boldsymbol{\theta} \circ \mathbf{p})). \quad (10)$$

In the simplified DistFlow equations above, the real/reactive branch flows and node voltages are entirely functions of real power injection $\mathbf{p}$, substation per unit voltage $V_0$ and distribution grid topology. The DSO can implement DLA to determine $\mathbf{p}$, using (8) – (10) to set up grid constraints, which are explained below.

*A.1 Transformer Capacity:* Since the distribution transformer(s) at the substation node has limited capacity, the total apparent power that the DSO can draw from the wholesale market is bounded as,

$$\mathbf{p}^T \mathbf{Z_0} \mathbf{p} \leq S_0^2. \quad (11)$$

Here, the matrix $\mathbf{Z_0}$ is given by,

$$\mathbf{Z_0} = \mathbf{1}_A \mathbf{1}_A^T + \boldsymbol{\theta} \boldsymbol{\theta}^T. \quad (12)$$

*A.2 Line Limits:* The total apparent power at line $k$ cannot exceed its MVA limit $\bar{S}_k$ so that $P_k^2 + Q_k^2 \leq \bar{S}_k^2$. Defining the matrix $\mathbf{E}_k$ as a square matrix with the only non-zero entry of unity appearing at the $k^{\text{th}}$ row and the $k^{\text{th}}$ column, it is seen that,

$$\mathbf{P}^T \mathbf{E}_k \mathbf{P} + \mathbf{Q}^T \mathbf{E}_k \mathbf{Q} \leq \bar{S}_k^2.$$

Whence, using (8) and (9), the following line limit constraint is obtained,

$$\mathbf{p}^T \mathbf{Z}_k \mathbf{p} \leq \bar{S}_k^2, \qquad \forall k \in \mathcal{N}. \tag{13}$$

Each matrix $\mathbf{Z}_k$ above is given by,

$$\mathbf{Z}_k = \mathbf{A}^T \mathbf{D}^T \mathbf{E}_k \mathbf{D} \mathbf{A} + diag\boldsymbol{\theta} \mathbf{A}^T \mathbf{D}^T \mathbf{E}_k \mathbf{D} \mathbf{A} diag\boldsymbol{\theta}. \tag{14}$$

*A.3 Node Voltage Limits:* The voltage at node $k$ is given by (10), and expressed solely in terms of $\mathbf{p}$ and other grid parameters. The total voltage deviation at node $k$ must not exceed a numerical value of $\delta$ (typically 0.05), yielding the following constraint,

$$\mathbf{1}_N - \boldsymbol{\delta} \leq V \leq \mathbf{1}_N + \boldsymbol{\delta}. \tag{15}$$

Whence using (10) in (15), and upon further simplification the following bounds on the power vector $\mathbf{p}$ are obtained,

$$\underline{\mathbf{l}} \leq \mathbf{M}\mathbf{p} \leq \bar{\mathbf{l}}. \tag{16}$$

In the above expression, the lower and upper bounds are,

$$\underline{\mathbf{l}} = (V_0 - 1)\mathbf{1}_N - \boldsymbol{\delta}, \qquad \bar{\mathbf{l}} = (V_0 - 1)\mathbf{1}_N + \boldsymbol{\delta}.$$

The matrix $\mathbf{M}$ that shows the sensitivity of voltage deviation to power injection $\mathbf{p}$ is equal to,

$$\mathbf{M} = \mathbf{M}_P + \mathbf{M}_Q diag\boldsymbol{\theta}, \tag{17}$$

where, $\mathbf{M}_P = V_0^{-1} \mathbf{Ur} \circ \mathbf{DA}$, $\mathbf{M}_Q = V_0^{-1} \mathbf{Ux} \circ \mathbf{DA}$.

*A.4 Budget Balance:* As in [27], [13], [10], the market price at the substation node is coupled with the demand drawn by the grid from the wholesale market and is modeled as,

$$c_0 = c_0^b + \beta_0 \sum_k p_k. \tag{18}$$

Here, $c_0^b$ is the base demand price and $\beta_0$ is the elasticity coefficient that can be obtained using statistics of historical data of locational marginal price as explained in [27]. The subscript '0' appears in all variables to indicate association with the root node. Note that to model $c_0$, the DSO can add a reasonable fixed amount to $c_0^b$ to account for any grid usage tariff. The underlying mechanism remains weakly budget balanced despite this increment. In other words, $\mathbf{c}^T\mathbf{p} - c_0 \mathbf{1}_A^T \mathbf{p} \geq 0$. Replacing $c_0$ from (18) yields the following DSO budget balance constraint,

$$c_0^b \mathbf{1}_A^T \mathbf{p} - \mathbf{c}^T \mathbf{p} + \beta_0 \mathbf{p}^T \mathbf{1}_A \mathbf{1}_A^T \mathbf{p} \leq 0. \tag{19}$$

### B. DSO Level Auction

*B.1 Social Welfare Problem:* The feasible set $\mathfrak{G}$ consists of all power vectors $\mathbf{p}$ that meet constraints in (11), (13), (16) and (19) pertaining to substation transformer MVA limit, lines MVA limit, nodes voltage limits, and DSO budget balance, i.e.

$$\mathfrak{G} = \left\{ \mathbf{p} \middle| \begin{array}{l} \forall k \in \mathcal{N} \cup \{0\}: \mathbf{p}^T \mathbf{Z}_k \mathbf{p} \leq \bar{S}_k^2, \\ \forall k \in \mathcal{N}: \underline{\mathbf{l}} \leq \mathbf{M}\mathbf{p} \leq \bar{\mathbf{l}}, \\ c_0^b \mathbf{1}_A^T \mathbf{p} - \mathbf{c}^T \mathbf{p} + \beta_0 \mathbf{p}^T \mathbf{1}_A \mathbf{1}_A^T \mathbf{p} \leq 0 \end{array} \right\}. \tag{20}$$

The goal of DSO is to solve the SW problem as stated below.

Maximize w.r.t. $[\mathbf{d}^k]_{k \in \mathcal{A}}, [\mathbf{s}^k]_{k \in \mathcal{A}}, \mathbf{p}$:

$$\Omega([\mathbf{d}^k]_{k \in \mathcal{A}}, [\mathbf{s}^k]_{k \in \mathcal{A}}, \mathbf{p}) = \sum_{k \in \mathcal{A}} \Theta_k(\mathbf{d}^k, \mathbf{s}^k, \mathbf{p}). \tag{21}$$

Subject to:

$$\mathbf{p} \in \mathfrak{G},$$
$$\mathbf{p} = \left[ \mathbf{1}_{N_B^k}^T \mathbf{d}^k - \mathbf{1}_{N_S^k}^T \mathbf{s}^k \right]_{k \in \mathcal{A}}, \tag{22}$$
$$\mathbf{s}^k \leq \mathbf{g}^k. \tag{23}$$

In this constrained optimization problem formulation, $\Omega$ is the global SW, obtained by aggregating that of each aggregator $\Theta_k$. The equality constraint in (22) restricts the quantity $p_k$ allocated to each aggregator k as the difference between the total energy demanded by the buyers ($\mathbf{1}_{N_B^k}^T \mathbf{d}^k$) and that supplied by the sellers ($\mathbf{1}_{N_S^k}^T \mathbf{s}^k$). The inequality constraint in (23) is present so that the power supplied $s_j^k$, by any seller $j$ stays below its generation $g_j^k$.

Each aggregator level SW $\Theta_k$ in (21) is given by,

$$\Theta_k(\mathbf{d}^k, \mathbf{s}^k | p_k) = \mathbf{1}_{N_B^k}^T \mathbf{u}^k + \mathbf{1}_{N_S^k}^T \mathbf{v}^k, \tag{24}$$

where $\mathbf{u}^k$ and $\mathbf{v}^k$ are the hidden utility functions of buyers and sellers in aggregator $k$.

Under the assumption that the utility functions $\mathbf{u}^k$ and $\mathbf{v}^k$ are concave, the SW optimization problem is convex. Note that in the DSO SW problem, $\mathbf{d}^k$ and $\mathbf{s}^k$ are local variable vectors pertaining to aggregator $k$'s buyers and sellers demand and supply whereas $\mathbf{p}$ is the global variable vector of injections to all aggregators. Due to lack of access to the utility functions $u^k, v^k$, and for scalability [28], [29], the DSO implements a distributed algorithm by decomposing its original problem into a master and multiple sub-problems that are solved in a distributed fashion. The master problem is solved by DLA and the sub-problems are solved by ALAs locally.

*B.2 Distributed Implementation:* At the lower level, each aggregator $k$, implements the following sub-problem in parallel by means of ALA:

Maximize w.r.t. $\mathbf{d}^k, \mathbf{s}^k: \Theta_k(\mathbf{d}^k, \mathbf{s}^k, \mathbf{p})$.

Subject to (22) and (23).

At the upper level, each iteration of DLA realizes a projected gradient descent step of the decomposed SW problem with $\mathbf{p}$ as the global variable. The DSO sends to each aggregator $k$ the energy $p_k$ and receives the gradient direction $\boldsymbol{\lambda} = \nabla_\mathbf{p} \Omega$ from it. It will be shown in Proposition 7 that $\boldsymbol{\lambda} = \mathbf{c}$ the vector of prices towards which ALA converges. Next, the vector $\mathbf{p}$ is incremented by an amount $\epsilon \boldsymbol{\lambda}$ to $\mathbf{p}''$, where $\epsilon$ is the gradient step size. The vector $\mathbf{p}''$ is then projected onto the feasible region $\mathfrak{G}$, after the constraint parameters are updated according to (14) and (17). Parameter updates are required because the reactive power changes according to the numerical value of the projection $\mathbf{p}'$, which in turn causes the fraction $\boldsymbol{\theta}$ to change.

The DLA algorithm terminates when any aggregator $k$ returns a flag $a_k = F$ defined in the following, indicating that the constraints in (22) and (23) were not satisfied,

$$a_k = \begin{cases} T \text{ if } p_k = \mathbf{1}_{N_B^k}^T \mathbf{d}^k - \mathbf{1}_{N_S^k}^T \mathbf{s}^k, \\ F \text{ if } p_k \neq \mathbf{1}_{N_B^k}^T \mathbf{d}^k - \mathbf{1}_{N_S^k}^T \mathbf{s}^k. \end{cases} \tag{25}$$

*Algorithm DLA:*

$\mathbf{p} \leftarrow$ initial
Repeat:
    Send to aggregators $k \in \mathcal{A}$: $p_k$
    Receive from aggregators $k \in \mathcal{A}$: $\lambda_k = c_k, a_k$
    If $\wedge_k a_k \neq T$ then exit loop
    $\mathbf{p}'' \leftarrow \mathbf{p} + \epsilon \lambda$
    Compute constraint parameters: $\boldsymbol{\theta}, \mathbf{M}, \mathbf{Z}_k$
    $\mathbf{p}' \leftarrow \underset{\mathbf{p} \in \mathfrak{G}}{\text{argmin}} \|\mathbf{p} - \mathbf{p}''\|$
Until convergence
Output: $\mathbf{p}, \mathbf{c}, [\mathbf{d}^k]_{k \in \mathcal{A}}, [\mathbf{s}^k]_{k \in \mathcal{A}}$

### C. Aggregator Level Auction

*C.1 Virtual Bidding:* Price anticipation is an undesirable effect that occurs in proportional auctions with relatively few bidders [3], [30]. In such scenarios, the bidders are aware of the sensitivity of the equilibrium price, i.e. that $\frac{\partial c_k}{\partial b_i^k} \neq 0$ for a buyer, and $\frac{\partial c_k}{\partial s_j^k} \neq 0$ for a seller. As the bidders place bids to maximize their individual payoffs, price anticipation leads to inefficiency. In [3], virtual bidding is shown to approach price-taking conditions for isolated microgrids by lowering the market powers of the bidders. Virtual bidding involves the presence of a virtual agent, which acts as both a seller and a buyer. Unlike other agents, the virtual bidder is merely an algorithmic entity that is incorporated within the aggregator, ergo has access to $\mathbf{b}^k$ and $\mathbf{s}^k$. By incorporating the virtual bidder, whose buying and selling bids can be as large as possible, the aggregator enables the price approach that of a market with potentially infinite number of agents. The larger the number of bidder agents, the less the effect of price anticipation [3]. Before addressing the implementation of ALA, the following propositions are established.

*Proposition–1:* Due to virtual bidding, ALA can be arbitrarily close to price taking mechanism. In other words, the following expressions hold,

$$c_k = \frac{\mathbf{1}_{N_B^k}^T \mathbf{b}^k}{p_k + \mathbf{1}_{N_S^k}^T \mathbf{s}^k}, \tag{26}$$

$$\frac{\partial c_k}{\partial b_i^k} = 0, \frac{\partial c_k}{\partial s_j^k} = 0. \tag{27}$$

*Proof:* The virtual bidder bids a large amount of energy $s_0$, which it buys for a total amount $c_k^0 s_0$. Here $c_k^0 = \frac{\mathbf{1}_{N_B^k}^T \mathbf{b}^k}{p_k + \mathbf{1}_{N_S^k}^T \mathbf{s}^k}$ is the desired price under price taking. The actual price, $c_k = \frac{c_k^0 s_0 + \mathbf{1}_{N_B^k}^T \mathbf{b}^k}{p_k + s_0 + \mathbf{1}_{N_S^k}^T \mathbf{s}^k}$, is computed by the aggregator. It can be readily established that $\lim_{s_0 \to \infty} c_k = \lim_{s_0 \to \infty} \frac{c_k^0 s_0 + \mathbf{1}_{N_B^k}^T \mathbf{b}^k}{p_k + s_0 + \mathbf{1}_{N_S^k}^T \mathbf{s}^k}$, justifying (26).

Likewise $\lim_{s_0 \to \infty} \frac{\partial c_k}{\partial b_i^k} = \lim_{s_0 \to \infty} \frac{1}{p_k + s_0 + \mathbf{1}_{N_S^k}^T \mathbf{s}^k} \left(1 + \frac{\partial c_k^0}{\partial b_i^k}\right) = 0$ and

$\lim_{s_0 \to \infty} \frac{\partial c_k}{\partial s_j^k} = \lim_{s_0 \to \infty} \frac{1}{p_k + s_0 + \mathbf{1}_{N_S^k}^T \mathbf{s}^k} \left(1 - \frac{c_k^0 s_0 + \mathbf{1}_{N_B^k}^T \mathbf{b}^k - \mathbf{1}_{N_S^k}^T \mathbf{s}^k}{p_k + s_0 + \mathbf{1}_{N_S^k}^T \mathbf{s}^k}\right) \frac{\partial c_k^0}{\partial s_j^k} = 0$,

so that (27) is valid in the limiting scenario. In other words, the presence of the virtual bidder that place very high bids allows the auction to behave as one with a large number of agents. Under these circumstances, the participating agents bid as price takers.

*Proposition-2:* ALA is strongly budget and energy balanced.

*Proof:* It follows from virtual bidding (26) that $c_k p_k + c_k \mathbf{1}_{N_S^k}^T \mathbf{s}^k = \mathbf{1}_{N_B^k}^T \mathbf{b}^k$. The RHS is the total monetary amount that the aggregator receives from the buyers. The LHS is the sum of the payment that the aggregator makes to the DSO and the reimbursement amount given to the sellers. This establishes strong budget balance. Energy balance is established under the proportional allocation auction paradigm [31], where the energy to each buyer to be proportional to its bid, i.e. $\mathbf{d}^k = \frac{1}{c_k} \mathbf{b}^k$. Energy balance immediately follows from (26). Note that this satisfies the constraint in in (22).

*C.2 Distributed Implementation:* ALA receives $p_k$ from the DSO, and initializes the price $c_k$ (see Fig. 2). In each step, it sends $c_k$ to the sellers and receives $\mathbf{s}^k$. Using proportional allocation, it determines $\mathbf{d}^k$ which is communicated to the buyers. The buyers return their bids $\mathbf{b}^k$. The price $c_k$ is determined as a two-step procedure using virtual bidding.

*Algorithm ALA(k):*

Receive from DSO: $p_k$
Initialize: $c_k$
Repeat:
    Send to sellers: $c_k$
    Receive from sellers: $\mathbf{s}^k$
    $\mathbf{d}^k \leftarrow \frac{1}{c_k} \mathbf{b}^k$
    Send to buyers: $\mathbf{d}^k$
    Receive from buyers: $\mathbf{b}^k$
    $c_k^0 \leftarrow \frac{\mathbf{1}_{N_B^k}^T \mathbf{b}^k}{p_k + \mathbf{1}_{N_S^k}^T \mathbf{s}^k}$
    $c_k \leftarrow \frac{c_k^0 s_0 + \mathbf{1}_{N_B^k}^T \mathbf{b}^k}{p_k + s_0 + \mathbf{1}_{N_S^k}^T \mathbf{s}^k}$
Until equilibrium
Send to DSO: $c_k$

### D. Equilibrium Analysis

*D.1 Bidding Strategies:* Buyers and sellers' bidding strategies are established by means of the following propositions.

*Proposition-3:* The bid $b_i^k$ placed by each buyer $i$ is such that its marginal utility equals its per unit price,

$$\frac{\partial u_i^k}{\partial d_i^k} = c_k. \tag{28}$$

*Proof:* The buyer's payoff $u_i^k(d_i^k) - b_i^k$, is maximized when its derivative is zero, i.e. $\frac{\partial}{\partial b_i^k} u_i^k(d_i^k) - 1 = \frac{\partial u_i^k}{\partial d_i^k} \frac{1}{c_k}\left(1 - \frac{1}{c_k}\frac{\partial c_k}{\partial b_i^k}\right) - 1 = 0$. Under virtual bidding, (27) holds so that (28) is satisfied.

*Proposition-4:* The bid $s_j^k$ placed by each seller $j$ is such that if the seller does not bid its entire generation ($s_j^k < g_j^k$), its marginal utility, with $\gamma_j^k \geq 0$ being a positive scalar quantity, equals its per unit price,

$$\frac{\partial v_j^k}{\partial s_j^k} = \gamma_j^k - c_k,$$
$$\begin{cases} \gamma_j^k = 0, & s_j^k < g_j^k, \\ \gamma_j^k > 0, & s_j^k = g_j^k. \end{cases} \quad (29)$$

*Proof:* In fact, $\gamma_j^k$ is a dual variable as shall be seen here. The seller's payoff $v_j^k(g_j^k - s_j^k) + c_k s_j^k$ is maximized under the constraint $s_j^k \leq g_j^k$. The Lagrangian for this problem is $L_j(s_j^k) = v_j^k(g_j^k - s_j^k) + c_k s_j^k - \gamma_j^k(s_j^k - g_j^k)$. At stationarity, $\frac{\partial}{\partial s_j^k} L_j = \frac{\partial}{\partial s_j^k} v_j^k(g_j^k - s_j^k) + c_k + s_j^k \frac{\partial c_k}{\partial s_j^k} - \gamma_j^k = 0$. Under virtual bidding, (27) holds so that (29) is satisfied.

*D.2 Aggregator Equilibrium:* During each iteration of DLA, an aggregator establishes equilibrium conditions to return price $c_k$.

*Proposition-5:* The equilibrium of the aggregator $k$'s auction maximizes the social welfare $\Theta_k(\mathbf{d}^k, \mathbf{s}^k, \mathbf{p})$ as defined in Eqn. (24) with respect to $\mathbf{d}^k, \mathbf{s}^k$ under the energy balance constraint, $p_k = \mathbf{1}_{N_B^k}^T \mathbf{d}^k - \mathbf{1}_{N_S^k}^T \mathbf{s}^k$ and with no seller selling more energy than its generated capacity, $\mathbf{s}^k \leq \mathbf{g}^k$.

*Proof:* The statement above defines a constrained optimization problem with the following Lagrangian $\mathcal{L}_k(\mathbf{d}^k, \mathbf{s}^k)$,

$$\mathcal{L}_k(\mathbf{d}^k, \mathbf{s}^k) = \Theta_k(\mathbf{d}^k, \mathbf{s}^k) - \boldsymbol{\gamma}_k^T(\mathbf{s}^k - \mathbf{g}^k)$$
$$- \lambda_k \left(\mathbf{1}_{N_B^k}^T \mathbf{d}^k - \mathbf{1}_{N_S^k}^T \mathbf{s}^k - p_k\right). \quad (30)$$

The constrained optimum satisfies ALA's energy balance condition. Stationarity at the optimum implies that $\nabla_{\mathbf{d}^k} \mathcal{L}_k = 0$ and $\nabla_{\mathbf{s}^k} \mathcal{L}_k = 0$. From (24), it is that $\nabla_{\mathbf{d}^k} \mathbf{u}^k = \lambda_k \mathbf{1}_{N_B^k}$ and $\nabla_{\mathbf{s}^k} \mathbf{v}^k = \boldsymbol{\gamma}_k - \lambda_k \mathbf{1}_{N_S^k}$, so that the optimum of $\Theta_k$ coincides with the auction equilibrium when $\lambda_k = c_k$ in (28) and (29) and with $\boldsymbol{\gamma}_k$ being the vector of entries $\gamma_j^k$ in (29).

*Proposition-6:* The price vector $\mathbf{c}$ returned by ALA is also the gradient of the overall SW function given by (21), i.e.

$$\nabla_\mathbf{p} \Omega([\mathbf{d}^k]_{k \in \mathcal{A}}, [\mathbf{s}^k]_{k \in \mathcal{A}}, \mathbf{p}) = \mathbf{c}. \quad (31)$$

*Proof:* From (28) and (29), it is seen that $\lambda_k = c_k$. It follows that $\frac{\partial}{\partial p_k}\Theta_k(\mathbf{d}^k, \mathbf{s}^k) = c_k$. The expression in (31) directly follows from (21).

*D.3 Global Equilibrium:*

*Proposition–7:* Under certain assumptions, the bilevel auction mechanism converges to the unique global maximum of the social welfare.

*Proof:* The matrix $\mathbf{Z}_0$ in (12) can be recognized to be symmetric positive semidefinite (SPSD). The matrix $\mathbf{E}_k$ in (14) is also an SPSD matrix with a single non-zero eigenvalue of unity. The first and second terms in the expression for $\mathbf{Z}_k$ in (14) being similar to $\mathbf{E}_k$ under the transform matrices $\mathbf{DA}$ and $\mathbf{DA}diag\boldsymbol{\theta}$, have the same eigenvalues and are therefore SPSD. Each matrix $\mathbf{Z}_k$ ($k \in \mathcal{N}$) being the sum of two such SPSD matrices is also SPSD. Hence, suppose $\mathbf{c}$ were constant, the region $\mathfrak{G}$ would be convex in $\mathbf{p}$ since it would contain only convex quadratic and linear inequality constraints.

However, $\mathbf{c}$ is dependent on $\mathbf{p}$ as the equilibrium price $c_k$ of any aggregator $k$ is determined by the power $p_k$ and the market mechanism. From (21) and (31), $\frac{\partial \Theta_k}{\partial p_k} = c_k$ whereas from (28) and (29), $\frac{\partial u_i^k}{\partial d_i^k} = -\frac{\partial v_j^k}{\partial s_j^k} = c_k$. Hence $\frac{\partial u_i^k}{\partial p_k} = \frac{\partial v_j^k}{\partial p_k} = c_k$. We assume that the utilities are strictly concave and increasing, i.e. that the marginal utility of the energy consumed is always positive no matter how small. Under these circumstances, neglecting cubic and higher order terms, the SW can be approximated as, $\Theta_k = c_k p_k + h_k p_k^2$, where $h_k < 0$. In fact, some recent research have modeled utilities as quadratic functions [32], [33], [18]. Thus $\Omega \approx \mathbf{c}^T \mathbf{p} + \mathbf{p}^T \mathbf{H} \mathbf{p}$, where $\mathbf{H}$ is a negative definite diagonal matrix thereby rendering $\Omega$ as strictly concave in $\mathbf{p}$. As the feasible region is strictly convex, termination of DLA due to a constraint violation can take place only when the constrained optimum has been reached with the violating constraint being active. DLA is a straightforward implementation of projected gradient ascent of the SW problem, whose convergence to the global minimum has been well studied [29], [34], [35].

*Proposition–8:* The bilevel auction is weakly budget balanced.
*Proof:* From Proposition-2, ALA is strongly budget balanced, so that $c_k \mathbf{1}_{N_B^k}^T \mathbf{d}^k - c_k \mathbf{1}_{N_S^k}^T \mathbf{s}^k = c_k p_k$. (19) explicitly imposes the weak budget balance constraint at the DSO level so that $\mathbf{c}^T \left[\mathbf{1}_{N_B^k}^T \mathbf{d}^k - \mathbf{1}_{N_S^k}^T \mathbf{s}^k\right]_{k \in \mathcal{A}} \geq c_0^b \mathbf{1}_A^T \mathbf{p} + \beta_0 \mathbf{p}^T \mathbf{1}_A \mathbf{1}_A^T \mathbf{p}$. The left side of this inequality is the total revenue collected from the sum total of all agents and the right side is the total payment made to the wholesale market. Since the latter can never exceed the money collected, the bilevel auction is weakly budget balanced.

III. SIMULATION RESULTS

Simulation results corroborates the theory presented in section II. A modified IEEE 37 node system as shown in Fig. 3 has been used to simulate the proposed bilevel energy allocation mechanism. A total of 483 agents, 303 buyers and 180 sellers, were generated and assigned to different nodes with aggregators. As summarized in Table I, seventeen aggregators (labeled A$k$) with different numbers of buyer and seller agents each with their own sets of parameters were created and assigned to nodes that have load. Fig. 3. shows a schematic of the IEEE 37 node system, where the nodes with aggregators are enclosed within shaded circles.

Due to their wide use in the literature for quantifying user satisfaction with diminishing returns, the utility functions of the agents were assumed to follow concave logarithmic curves [36]-[38] according to the equations $u_i(d_i) = x_i \log(y_i d_i + 1)$ and $v_j(g_j - s_j) = x_j \log(y_j(g_j - s_j) + 1)$. The quantities $x_i, y_i, x_j$ and $y_j$ in these equations were generated randomly, separately for each agent. They were adjusted so that agents' marginal utilities are scaled to reasonable per unit prices. The generation $g_j$ for sellers were also drawn at random, uniformly in the interval [0.1, 0.5] pu based on a system base value of 100 kVA.

TABLE I
AGGREGATOR ASSIGNMENT WITH THEIR RESPECTIVE NUMBER OF BUYER AND SELLER AGENTS TO NODES

| $Ak$ | 1 | 2 | 3 | 4 | 5 | 6 | 7 | 8 | 9 | 10 | 11 | 12 | 13 | 14 | 15 | 16 | 17 |
|---|---|---|---|---|---|---|---|---|---|---|---|---|---|---|---|---|---|
| Node | 1 | 8 | 12 | 13 | 17 | 18 | 22 | 23 | 25 | 26 | 27 | 29 | 30 | 31 | 33 | 35 | 36 |
| $N_B^k$ | 14 | 11 | 25 | 27 | 20 | 8 | 21 | 27 | 22 | 21 | 5 | 13 | 7 | 19 | 10 | 26 | 27 |
| $N_S^k$ | 13 | 7 | 5 | 21 | 22 | 10 | 15 | 6 | 4 | 22 | 3 | 9 | 4 | 19 | 5 | 6 | 9 |

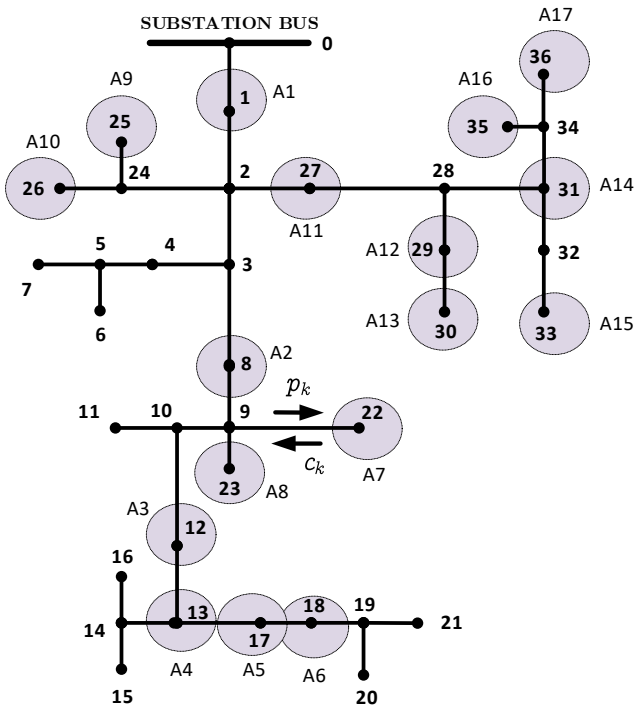

Fig. 3. Modified IEEE 37 node system with aggregators indexed. $A_k$ is used as an abbreviation for aggregator $k$.

Four different scenarios were created to examine the impact that the price and substation node's capacity have on DLA's allocation of energy to each aggregator and ALA allocations to local buyer and seller agents along with the resulting global SW. These four scenarios were created with differing wholesale market price and substation transformer capacity parameters, according to (18) and (11) as follow.

I) $c_0^b = 800¢/\text{pu}$ (high), $\beta_0 = 40¢/\text{pu}^2$ (elastic), $S_0 = 25$ pu
II) $c_0^b = 200¢/\text{pu}$ (low), $\beta_0 = 30¢/\text{pu}^2$ (elastic), $S_0 = 25$ pu
III) $c_0^b = 200¢/\text{pu}$ (low), $\beta_0 = 10¢/\text{pu}^2$ (elastic), $S_0 = 25$ pu
IV) $c_0^b = 200¢/\text{pu}$ (low), $\beta_0 = 0¢/\text{pu}^2$ (inelastic), $S_0 = 40$ pu

In Scenario I, in order to draw minimum energy from the substation bus and observe energy trade among aggregators, the base price $c_0^b$ and demand sensitivity coefficient $\beta_0$ were increased significantly to $800¢/\text{pu}$ and $40¢/\text{pu}^2$. In Scenarios II and III, equal base prices but different sensitivity coefficients were selected to simulate two normal scenarios. In Scenario IV, a base price of $200¢/\text{pu}$ with zero sensitivity coefficient was selected so that the price is inelastic to demand to observe energy trade and allocation under cheap supply from the wholesale market. Furthermore, the transformer capacity was increased to 40 pu to draw high amount of energy from the wholesale market and observe activation of physical grid constraints in DLA.

All case studies were simulated using MATLAB. Fig. 4 shows the outcome of the bilevel mechanism for each scenario. Here bars show the energy allocation $p_k$ to each node that has aggregator and dotted lines show its corresponding price. The energy allocation $p_k$ to each aggregator $k$ (shown in Fig. 3 as an $Ak$) and the equilibrium price $c_k$ are the solutions of DLA and ALA at equilibrium when the global SW given by (21) has been reached its constrained maximum. Furthermore, the market equilibrium price $c_k$ are different for each aggregator, illustrating that DLA is price heterogeneous.

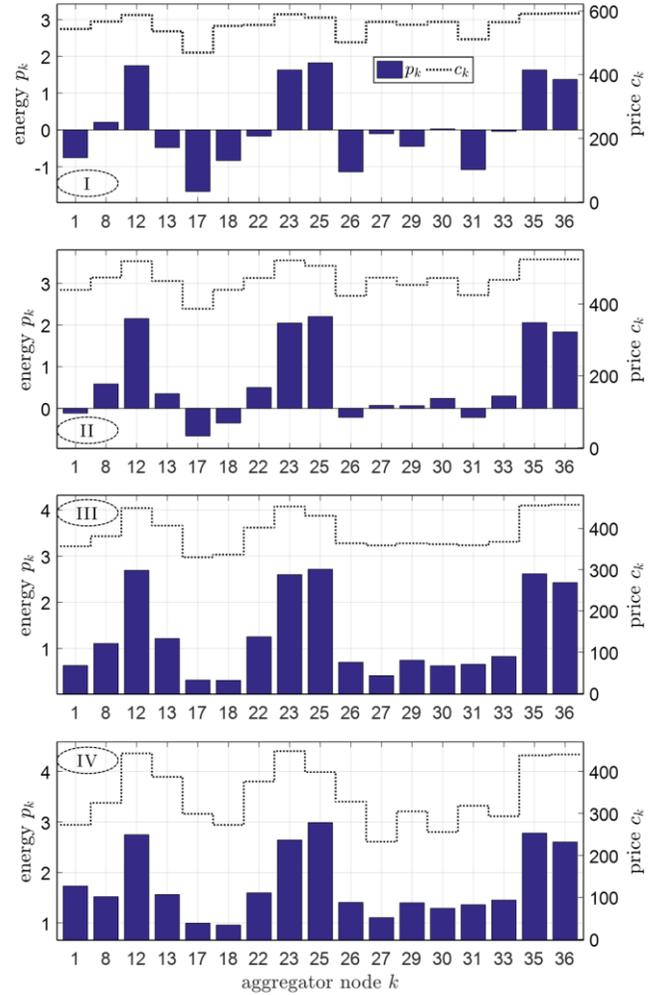

Fig. 4. DLA outcome for aggregators' share of energy (in pu) and its associated price (in $¢/pu$)

In Scenario I, due to significantly high wholesale price, the aggregators are willing to trade among themselves at different prices ranging from 469 ¢/pu to 592 ¢/pu. Aggregators with more buyers, i.e. A3 on node 12 ($N_B^3 = 25$, $N_S^3 = 5$), buy more. Several aggregators, especially those with more sellers, sell energy. For example, A5 on node 17 supplies a high amount of energy as it has more sellers ($N_S^5 = 22$), compared to others. The DSO draws only 1.72 pu at $c_0$ of 869 ¢/pu from the wholesale market at which it becomes strongly budget balanced, i.e. the constraint in (19) is active and the DSO makes zero profit.

In Scenario II, the price $c_0$ decreases to 525 ¢/pu. As a result, a smaller number of aggregators sell energy to the DSO. The ALA equilibrium prices $c_k$ in this scenario range from 387 ¢/pu to 524 ¢/pu. The DSO draws 10.83 pu from the wholesale market making zero profit. Aggregators import more energy compared to Scenario I. For instance, A4 on node 13 that was supplying energy as a seller in Scenario I imports energy as a buyer in Scenario II.

In Scenario III, the wholesale market price drops down to 419 ¢/pu at which point, all aggregators switch to being buyers. At this point, the DSO draws 21.87 pu from the wholesale market while still not attaining any budget surplus.

In Scenario IV, the wholesale market price at 200 ¢/pu is inelastic and inexpensive. All aggregators import energy at different equilibrium prices ranging from 233 ¢/pu to 448 ¢/pu. Aggregators with more buyers and with steeper utilities, such as those in nodes 12, 25, and 35, import high amounts. The DSO in this Scenario draws 30.14 pu and makes 4992 cents in the role of an arbitrager. It is worth emphasizing that the proposed mechanism is an SW maximizing bilevel auction in which the goal is to maximize the overall SW. The DSO's budget is merely set as a constraint. As pointed earlier, a profit seeking DSO can incorporate its surcharge in the pricing model in (18).

Note that the volume of energy traded at each aggregator node depends internally on the numbers of buyer and seller agents ($N_S^k$ and $N_B^k$), generation capacities $g_j^k$, and marginal utilities ($u_i'^k$ and $v_j'^k$) as well as all external factors such as the wholesale price $c_0$ and other aggregators' market conditions. In general, aggregators with a surplus of energy that settle down to lower equilibrium prices $c_k$ tend to supply more $p_k$ to the rest of the network, whereas those that converge to higher $c_k$ supply less. Similarly, aggregators with deficit energy and higher equilibrium prices $c_k$, are assigned more $p_k$.

Convergence to the global optimum of the SW (sum of utilities of 483 agents) under each scenario is shown in Fig. 5. As can be seen, in all cases, DLA converges to within 1% of the optimum within as little as 10 iterations. The maximum allowable iterations of ALA to converge to equilibrium is 100 per DLA iteration in all cases, although it was routinely observed that equilibrium could be established well before this limit.

In Fig. 5, it can be observed that as the injection $\mathbf{1}_A^T \mathbf{p}$ from the wholesale market increases, so does the SW. This is because the availability of more energy to the buyers results in their acquiring less from sellers. Consequently, selling agents consume a larger portion of their own generated energy.

In Fig. 6, the node voltages $V_k$ are shown, sorted according to the aggregator index $k$, separately for each scenario. The upper and lower bounds on the pu voltages appear as horizontal

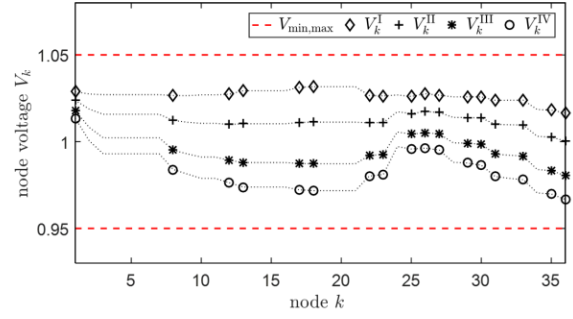

Fig. 6. Node pu voltages within given bounds.

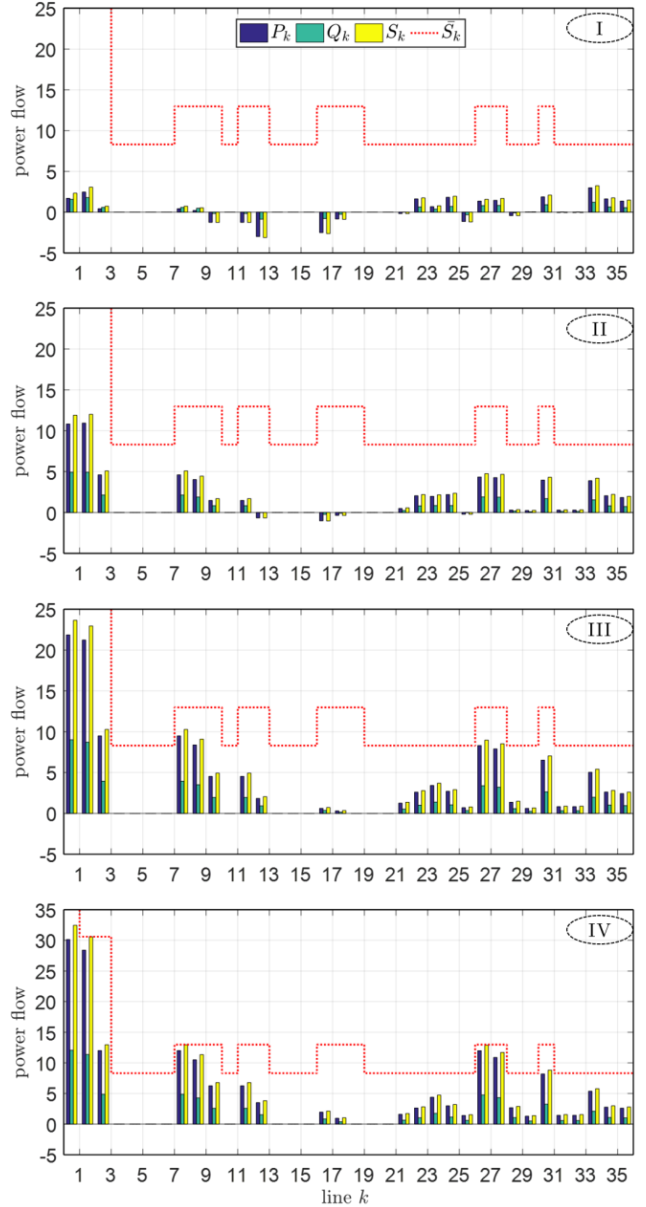

Fig. 7. Real, reactive, and apparent power flows in line $k$ (line entering node $k$) within line MVA limits.

dotted lines. In all cases, it can be seen that the voltages remain within these bounds, indicating that the voltage constraint in (15) is met. Increasingly higher voltage drops can be seen from

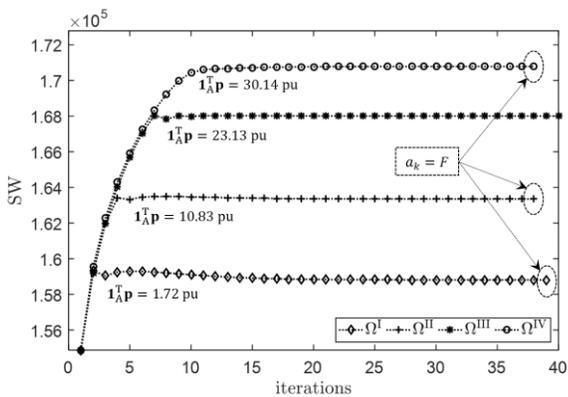

Fig. 5. Optimum SW under each scenario. In scenarios other than III, some aggregator $k$ returns $a_k = F$ and DLA terminates.

Scenarios I through IV. This is a direct outcome of increasing supply from the wholesale market.

Line power flows are shown in Fig. 7. The dotted lines pertain to the line MVA limits ($\bar{S}_k$). Adjacent vertical bars correspond to real ($P_k$), reactive ($Q_k$), and apparent ($S_k$) power flows of each line segment, which are again ordered according to the indices of their corresponding nodes. Higher energy supplied from the wholesale market results in more power flows (top – bottom). Moreover, in Scenario I (top), the power flows in several lines are in the reverse direction, as it appear below the x-axis. Conversely, in Scenario IV (bottom), all lines that have downstream aggregators can be seen to be delivering power to them. Additionally, the powers flowing through three lines (indexed 2, 8, 27) are at their limits (13). The adjacent set of lines labeled 12, 13 and 17 as per their node indices (see Fig. 3) which provide power directly to aggregators A3, A4, and A5 are taken up as examples to further highlight some observations.

In Scenario I, due to significantly high wholesale price $c_0$, with high number of sellers and low price, A4 and A5 on nodes 13 and 17 supply to A3 on node 12 that has 25 buyers and only 5 sellers (see Fig. 4 and Fig. 7). In Scenario II, however, as $c_0$ is decreased, A4 on node 13 starts to import (see Fig. 4) while A5 keeps exporting to feed A3 as well as A4. This forces power flow in lines indexed 17 and 13 to become negative regardless of the fact that A4 imports. Notice also that A3 imports from other upstream aggregators in addition to A5 resulting to a positive power flow in line 12. In Scenarios III and IV, due to further decrease in the wholesale price $c_0$, all aggregators import power which result to positive power flow in all lines.

So far, we discussed results pertaining to DLA, distribution grid constraints, and power flows. In order to illustrate the results of ALAs, auction outcome for A6 ($N_B^6 = 8$, $N_S^6 = 10$) under Scenarios I and IV are summarized in Table II. In Scenario I, A6 exports 0.834 pu to the grid which is obtained as the sum of supplies ($\sum_j s_j^6$) of its sellers minus sum of demands of its buyers ($\sum_i d_i^6$). The price $c_6$ stabilizes at 533 ¢/pu at which only those buyers and sellers' whose marginal utilities $u_i'^6$ and $v_j'^6$ becomes equal to this price can trade. Furthermore, because of high wholesale price $c_0$, all sellers are willing to trade and declare nonzero $s_j^6$. For example, seller 1 generates $g_1^6 = 0.434$ pu and sells $s_1^6 = 0.325$ pu at $v_j'^6 = c_0 = 553$.

In Scenario IV, A6 imports 0.958 pu at a lower equilibrium price of 273 ¢/pu. Buyers' demand increase and only sellers with marginal utilities $v_j'^6$ equal to $c_6$ trade nonzero supply $s_j^6$. Notice that sellers 3, 6, 7, and 8 whose marginal utilities are higher than $c_6$, i.e. $v_j'^6 > c_6$, are assigned zeros supply $s_j^6$, allowing them to consume all their generation $g_j^6$ instead of selling so that their utilities are maximized. Notice that the marginal utilities of all agents (both buyers and sellers) that are trading becomes uniform and equal to the market equilibrium price $c_6$ for both scenarios.

TABLE II
ALA OUTCOME FOR AGGREGATOR 6 (A6) UNDER SCENARIOS I AND IV

| Aggregator 6 on node 18 ($\mathcal{N}_B^6 = 5, \mathcal{N}_S^6 = 20$) | | | | | |
|---|---|---|---|---|---|
| $p_6$ (pu) | | $c_6$ (¢/pu) | | $V_6$ (pu) | |
| I | IV | I | IV | I | IV |
| -0.834 | 0.958 | 553 | 273 | 1.032 | 0.972 |
| Buyer Agents | | | | | |
| $d_i^6$ (pu) | | $c_i^6$ (¢/pu) | | $u_i'^6$ (¢/pu) | |
| I | IV | I | IV | I | IV |
| 0.115 | 0.234 | 553 | 273 | 553 | 273 |
| 0.105 | 0.214 | 553 | 273 | 553 | 273 |
| 0.118 | 0.240 | 553 | 273 | 553 | 273 |
| 0.079 | 0.160 | 553 | 273 | 553 | 273 |
| 0.131 | 0.267 | 553 | 273 | 553 | 273 |
| 0.126 | 0.257 | 553 | 273 | 553 | 273 |
| 0.081 | 0.165 | 553 | 273 | 553 | 273 |
| 0.123 | 0.251 | 553 | 273 | 553 | 273 |
| Seller Agents | | | | | |
| $g_j^6$ (pu) | | $s_j^6$ (pu) | | $v_j'^6$ (¢/pu) | |
| I | IV | I | IV | I | IV |
| 0.434 | 0.434 | 0.325 | 0.211 | 553 | 273 |
| 0.437 | 0.437 | 0.351 | 0.262 | 553 | 273 |
| 0.110 | 0.110 | 0.017 | 0 | 553 | 467 |
| 0.232 | 0.232 | 0.151 | 0.067 | 553 | 273 |
| 0.270 | 0.270 | 0.141 | 0.007 | 553 | 273 |
| 0.153 | 0.153 | 0.018 | 0 | 553 | 490 |
| 0.249 | 0.249 | 0.127 | 0 | 553 | 272 |
| 0.185 | 0.185 | 0.086 | 0 | 553 | 295 |
| 0.274 | 0.274 | 0.182 | 0.087 | 553 | 273 |
| 0.432 | 0.432 | 0.315 | 0.194 | 553 | 273 |

## IV. CONCLUSION

A globally efficient energy allocation mechanism that does not violate any grid constraints has been proposed here. At the DSO level, the auction is price heterogeneous among aggregators, while the lower level auction is price uniform among home agents. Instead of determining the energy price as in other recently proposed mechanisms [8]-[10] the upper level agent (DSO) allocates power to the aggregators, that in turn conduct their separate ALA to establish market equilibrium conditions locally within their agents.

One of the advantages of this new approach is that the aggregators can conduct their own auctions in islanded conditions. During islanding, the aggregator would assume zero power allocated from the DSO. Its ALA would converge to an appropriate unit price entirely via market equilibrium, allowing locally energy trade among its agents to proceed.

The proposed mechanism is able to maximize SW requiring access to neither the agents' utility functions nor their power generations. ALA is able to do so by inferring the derivatives of the agents' utility functions from bidding information (see Proposition 5). This is another advantage of the proposed bilevel mechanism.

As ALA relies on market equilibrium prices, it is an iterative mechanism, requiring several iterations to establish equilibrium. As a tradeoff, DLA requires relatively few iterations to converge to the SW maximizing power allocation. This is because it uses primal problem decomposition instead of the dual decomposition approach. An algorithm to maximize SW that would use dual decomposition, would need gradient information, whereas unlike in primal decomposition, the derivatives of the agents' utilities with respect to allocated energy would not be readily available to the DSO. The alternative is for a dual decomposition optimization scheme to apply subgradient descent, which is known to converge very slowly (see [28] for details).

There are certain limitations of the research described below that can be addressed effectively in future research. The simplified DistFlow equations used here neglect power loss. The underlying physical grid model can be extended to consider line losses by using a more elaborate set of power flow equations [2].

The bilevel auction assumes a radial distribution system. This is because radial networks are very common, and prevalent within the United States. In order to adapt the proposed bilevel auction mechanism to other networks, only the manner in which the physical grid constraints are imposed (using DistFlow equations) would need to be replaced appropriately.

The present approach assumes a balanced network. The mechanism could be extended for use in 3-phase unbalanced networks by implementing additional grid constraints specifying the maximum allowable power imbalance, along with some algorithmic modifications.

Future research can also investigate how to extend the current model to multiple time slots. Aggregators that do not include any energy storage or conventional generators can implement an auction during each time slot independently of the others. For aggregators that do so, temporal constraints would need to be taken into account.